\documentclass[12pt,a4paper]{article}
\pdfoutput=1
\usepackage{myjheppub}
\usepackage{epstopdf}
\usepackage{graphicx}
\usepackage{epsfig}
\usepackage{dcolumn}  
\usepackage{bm}    
\usepackage{amssymb} 
\usepackage{amsmath,bm}
\usepackage{amsfonts}    
\usepackage{slashed}  
 \usepackage{relsize}
\usepackage[mathscr]{euscript}
\usepackage{epsfig}
\usepackage{color}

\def\nn{\nonumber}

\def\bz{{\bar{z}}}

\def\w32{\mathcal{W}_3}
\def\w32{\mathcal{W}_3^{(2)}}

\def\order{\mathcal{O}}
\def\tr{\text{tr}}
\def\S{\mathcal{S}}
\def\cL{\mathcal{L}}
\def\cW{\mathcal{W}}
\def\A{ \mathscr{A}  }
\def\B{ \mathscr{B}}
\def\c{\mathsf{c}}
\def\H{\langle H \rangle}
\def\dim{\textsf{dim}}

\title{Relative entropy in higher spin holography}

\author{Shouvik Datta} 
 
\affiliation{Centre for High Energy Physics, Indian Institute of Science,\\ C.~V.~Raman Avenue, Bangalore 560012, India.}

\emailAdd{shouvik@cts.iisc.ernet.in}

\abstract{We examine relative entropy in the context of the higher-spin/CFT duality. We consider 3$d$ bulk configurations in higher spin gravity which are dual to the vacuum and  a high temperature state of a CFT with $\cW$-algebra symmetries in presence of a chemical potential for a higher spin current. The relative entropy between these states is then evaluated using the Wilson line functional for holographic entanglement entropy.
In the limit of small entangling intervals, the relative entropy should vanish for a generic quantum system. We confirm this behaviour by showing that the difference in the expectation values of the modular Hamiltonian between the states matches with the difference in the entanglement entropy  in the short-distance regime. Additionally, we compute the relative entropy of states corresponding to smooth solutions in the $SL(2,\mathbb{Z})$ family with respect to the vacuum. }

\begin{document}

%
%
%
%

 \maketitle 

%

\section{Introduction}

A major theme in the context of the AdS/CFT correspondence \cite{Aharony:1999ti} has been the study of dualities between theories of higher spin gravity and CFTs with extended symmetries. Higher spin theories of gravity has considerably lesser fields than full-fledged string theories. At the same time, one hopes to capture some features of stringy geometries since they go beyond classical supergravity. The CFTs dual to these higher spin theories are not necessarily strongly coupled and are reasonably tractable.  This enables us to learn a lot from both sides of the duality and understand holography at a deeper level.

The first of such higher-spin/CFT dualities was the one which related the $O(N)$ vector model in 3 dimensions to Vasiliev gravity in $AdS_4$ \cite{Klebanov:2002ja,Sezgin:2002rt}. A lower dimensional version was proposed by Gaberdiel and Gopakumar which related $\cW_N$ minimal models (or WZW coset CFTs) to higher spin gravity in three dimensions \cite{Gaberdiel:2010pz,Gaberdiel:2012uj}. These dualities are essentially non-supersymmetric, unlike traditional versions of the AdS/CFT correspondence. There are now embeddings of both these dualities in string theory \cite{Gaberdiel:2013vva,Chang:2012kt}. 

Unlike Vasiliev theory in four spacetime dimensions, higher spin gravity in three dimensions admits a simpler description in terms of a Chern-Simons theory based on the gauge group hs$[\lambda]$. Also, 2$d$ CFTs are more tractable compared to higher dimensional ones. There has been a growing body of evidence in favour of the duality which includes matching of the spectrum, asymptotic symmetries and correlation functions \cite{Henneaux:2010xg, Campoleoni:2010zq,Gaberdiel:2011zw,Gaberdiel:2011wb,Hijano:2013fja}. 
Moreover, there exist explicit constructions for classical solutions in higher spin gravity in 3$d$ \cite{Gutperle:2011kf,Kraus:2011ds,Castro:2011iw,Campoleoni:2013lma,Bunster:2014mua}. The thermodynamics of the black holes has been extensively studied \cite{Perez:2012cf,David:2012iu,deBoer:2013gz,Chen:2012ba,Beccaria:2013dua,Compere:2013nba} and some of their properties (like partition functions, correlation functions etc.) have been strikingly matched from the dual CFT \cite{Gaberdiel:2012yb,Kraus:2011ds,Gaberdiel:2013jca} . See \cite{Ammon:2012wc,Perez:2014pya} for recent reviews on the subject. 

In the holographic context, another remarkable development is the evolution of geometrical methods for calculations of entanglement entropy in strongly-coupled field theories. Entanglement entropy is useful measure in quantum information theory, many-body physics and QFTs. It characterizes various topological phases, serves as an order parameter for phase transitions and has interesting connections to the entropy of black holes \cite{Osterloh,Amico:2007ag,Bombelli:1986rw,Holzhey:1994we,Calabrese:2004eu,Calabrese:2005in,Calabrese:2009qy,Srednicki:1993im}. 

Entanglement entropy is typically difficult to calculate in quantum field theories. The Ryu-Takayanagi prescription however offers a simple and elegant route for the calculation of holographic entanglement entropy \cite{Ryu:2006bv,Ryu:2006ef}. It states that entanglement entropy of a sub-system $A$, is given by the minimal surface in $AdS$ that ends on the boundary of $A$. This formula obeys strong subadditivity and has reproduced the area law in 2$d$-CFTs, amongst its other successes.  

Since higher spin gravity goes beyond diffeomorphism invariance, one needs to rethink usual notions of differential geometry -- like black hole singularities, horizons and the Ryu-Takayanagi minimal surfaces. Framed in the language of Chern-Simons theory, this means that we need to consider observables that are gauge invariant. The classical solutions which have been constructed have a gauge invariant characterization in terms of holonomies. The proposal for the holographic entanglement entropy for the case of higher spin gravity was given in terms of a Wilson line functional in \cite{deBoer:2013vca} and \cite{Ammon:2013hba}. We shall be using the proposal of \cite{deBoer:2013vca} for the holographic calculations in this paper. The calculations using the Wilson line from gravity was also recently matched with direct computations in $\cW$-algebra CFTs \cite{Datta:2014ska,Datta:2014uxa}. 

In this work, we shall explore the concept of relative entropy in the context of higher spin holography. The relative entropy is a measure which characterizes how different two quantum states are. If $\sigma$ and $\rho$ are density matrices corresponding to the states,  the relative entropy is defined as
\begin{align}
\cal  S(\sigma || \rho) = \tr \left(\sigma \log {\sigma} \right) -  \tr \left(\sigma \log {\rho} \right). \nn
\end{align} 
 There also exists a `thermodynamic-like' relation which relates the difference in entanglement entropy ($\Delta S$) and modular Hamiltonian $(\Delta \H)$ to relative entropy. Both these quantities are calculable using holographic techniques and one might expect to calculate the relative entropy as well \cite{Blanco:2013joa} (see \cite{Bhattacharya:2012mi,Lashkari:2013koa,Bhattacharya:2013bna,Faulkner:2013ica,Wong:2013gua,Banerjee:2014oaa,Lashkari:2014yva} for related works). We shall verify some of the key properties of relative entropy in $\cW$-algebra CFTs when deformed by a spin-3 chemical potential. The calculations will involve the Wilson line formula for holographic entanglement entropy. We shall evaluate the relative entropy between a high temperature state (dual to a higher spin black hole \cite{Gutperle:2011kf}) and the vacuum at zero temperature (dual to a generalization of the global-$AdS_3$ for non-zero chemical potentials \cite{Compere:2013nba}).

 \subsection*{Statement of the problem and summary of results} 
 The relative entropy between two states can be expressed as 
 \begin{align}  
 \S (\sigma ||\rho ) = \Delta \langle H \rangle- \Delta S \nn,
 \end{align}
 where $\Delta \langle H \rangle$ is the difference in the expectation values of the modular Hamiltonians between the states and $\Delta S$ is the corresponding variation in entanglement entropy.
 
 The relative entropy also obeys the following property\footnote{We denote the dimension of the Hilbert space  ($\mathcal{H}_A$) of subsystem $A$  by $\dim(A)$. $A'$ is the complement or the rest of the system.}  
 \begin{align}  
 &\lim_{ \left( \frac{\dim(A)}{\dim(A')}  \right) \rightarrow 0} \S( \tr_{A'} \sigma_A  || \tr_{A'} \rho_A ) =0 .\nn
 \end{align}
  That is, in the regime of small sub-system sizes or short entangling intervals {$\dim{(A)}\ll\dim{(A')}$}, we expect 
  \begin{align}
 \Delta \langle H \rangle = \Delta S  .\nn 
  \end{align}
It is this relation which we verify holographically for a large-$\c$ CFT with a $\cW_3$ symmetry and perturbed by a chemical potential for the spin-3 current.
 The relative entropy is calculated between an high-temperature state and the vacuum.
The bulk configuration corresponding to the vacuum is a generalization of the global-$AdS_3$ at finite chemical potential but still at zero temperature -- we shall be referring to this as the `higher spin vacuum' (see subsection 3.5 of \cite{Compere:2013nba}). On the other hand, at high temperatures the higher spin black hole \cite{Gutperle:2011kf} is the dominant saddle. We evaluate the holographic entanglement entropy corresponding to these solutions using the holomorphic Wilson line functional. The modular Hamiltonians are calculated from the holographic stress tensors. 

We consider a CFT with a central charge\footnote{Throughout the paper we use $\c$ as the notation for the central charge. This is because the other $c$, makes an appearance in the modular parameter $\frac{a\tau +b}{c\tau +d}$.} $\c$ on a circle of circumference $\ell\,(=2\pi R)$. We obtain that at leading order in the interval size $R\phi$, the expressions for $\Delta S$ and $\Delta \H$ match.
\begin{align}
&\Delta S \Big|_{\text{to }\phi^2} \ \bm{=} \ \Delta H \Big|_{\text{to }\phi^2} \nn \\ 
=&{ \c \phi^2\Bigg[ \frac{ \left((\ell T)^2 + 1\right)}{72} 
+\nn 
\frac{5\left((\ell T)^4-1\right)}{54}  \frac{\mu ^2}{ R^2}
+
\frac{20 \left((\ell T)^6+1\right)}{27} \frac{\mu^4}{ R^4}
+
\frac{1768  \left((\ell T)^8-1\right)}{243}  \frac{\mu^6}{ R^6} +... \Bigg]  }
\end{align}
Here, $T$ is the temperature of the high-temperature state and $\mu$ is the spin-3 chemical potential. 
We have been able to show this matching in a perturbative expansion in $(\mu/R)$. One can, in fact, keep track of terms to arbitrary orders. 
 
Furthermore, we calculate the relative entropy with respect to the vacuum in the $SL(2,\mathbb{Z})$ family of bulk solutions. The relative entropy may act an important measure  of quantum distinguishability of the states corresponding to these bulk configurations from the dual CFT viewpoint. The bulk solutions are modular images of $AdS_3$. They have a non-zero contribution to the full modular invariant partition function, $Z[\tau] = \sum Z_{AdS_3}\left[   \frac{a\tau +b }{c\tau +d} \right]$ \cite{Maldacena:1998bw,Dijkgraaf:2000fq,Manschot:2007ha,Li:2013rsa}. The result which we shall obtain is relevant even if one considers pure Einstein gravity (we need to set $\mu$ to zero).

 \subsection*{Outline of the paper}
 
 This paper is organized as follows. In section \ref{rel-e}, we review the definition of relative entropy, the modular Hamiltonian and its connections with the first law of entanglement. Section \ref{hs-ee} contains a review the Chern-Simons formulation of higher spin gravity and classical solutions constructed therein. We then utilize the Wilson line prescription to calculate the holographic entanglement entropy corresponding to these classical solutions. In section \ref{rel-calc}, we calculate the relative entropy between a high temperature state and the ground state using holographic methods. It is then shown that the relative entropy has the desired property in the short distance regime. We also calculate the relative entropy between a state corresponding to a smooth solution in the $SL(2,\mathbb{Z})$ family in the bulk and the vacuum. Section \ref{conclude} has our conclusions. 
 
\section{Relative entropy and the first law of entanglement}
\label{rel-e}
\subsection{Relative entropy : definition and properties}

In this section, we shall briefly review the definition and various properties of relative entropy which are relevant for the present problem. The reader may refer to \cite{Vedral:2002zz,nielsen2010quantum,Blanco:2013joa} for more details.  

The relative entropy is useful measure characterizing the closeness between two states in the same Hilbert space. For two density matrices $\sigma$ and $\rho$, the relative entropy can be defined as follows\footnote{To avoid confusion we shall be using $S$ for entanglement entropy and $\S$ for relative entropy.}
\begin{align} \label{re-def}
\cal  S(\sigma || \rho) = \tr \left(\sigma \log {\sigma} \right) -  \tr \left(\sigma \log {\rho} \right). 
\end{align} 
For example, if $\rho_{gr}$ is the density matrix corresponding to the ground state of the system and $\rho_{ex}$ being that of an excited state, then $\S(\rho_{ex} || \rho_{gr})$ gives a measure of `distinguishability' of the excited state from the ground state. Note that, \eqref{re-def} is not symmetric under the exchange of $\sigma$ and $\rho$ and doesn't strictly qualify as a measure of distance between two states. 

It can be shown that the relative entropy obeys the following properties
\begin{enumerate}
\item It is always non-negative (Klein's inequality)
$$\S(\sigma || \rho) \geq 0.$$
\item It is invariant under unitary transformations on the density matrices
$$
{\cal S}(\sigma || \rho) = \mathcal{S}(U^\dagger \sigma U ||U^\dagger  \rho U) . 
$$
Since, unitary transformations are just a change of basis, it should not affect the distinguishability between two states. 
\item It decreases upon partial tracing
$$
\S(\sigma || \rho) \geq \S( \tr_P \sigma  || \tr_P \rho ) . 
$$
This is true because the act of partial tracing restricts the amount information about the states and it becomes harder to distinguish between them. 
\item It is additive 
$$
{\cal S}(\sigma_A \otimes \sigma_B || \rho ) = \S(\sigma_A || \rho ) + \S(\sigma_B || \rho ) .
$$
This is allowed by the non-symmetric definition of relative entropy \eqref{re-def}.
\end{enumerate}

The properties 2 and 3 above imply a general class of maps which ensures the non-increasing feature of relative entropy \cite{Vedral:2002zz}. These are linear mappings which maintain the unit trace and hermiticity of the density matrix.  They are referred to as Completely Positive and Trace Preserving (CPTP) maps. 
\def\r{\mathscr{Y}}

The properties 1 and 3 also imply a monotonically decreasing property. For a pair of spatial bipartitionings of the system into $A\cup A'$ and $B\cup B'$ we have
\begin{align}
&\S ( \tr_{A'} \sigma_A  || \tr_{A'} \rho_A )  \geq \S  ( \tr_{B'} \sigma_B  || \tr_{B'} \rho_B )   \text{  \ for } A \supseteq B    .
\end{align}
%
Combining this with the non-negativity property of relative entropy, we have
\begin{align}  \label{small-size}
 \quad &\lim_{  \frac{\dim(A)}{\dim(A')}  \rightarrow 0^+} \S( \tr_{A'} \sigma_A  || \tr_{A'} \rho_A ) =0 .
 \end{align}

\subsection{Modular Hamiltonian and entanglement entropy}

For a reduced density matrix $\rho$, corresponding to a particular many-body system or quantum field theory, the modular Hamiltonian (or entanglement Hamiltonian)  $H$ is defined by the following relation
\begin{align}
\rho = \frac{e^{-H}}{\tr (e^{-H})}.
\end{align}
One is allowed to express the reduced density matrix in the above form since it is positive-semidefinite and Hermitian. 

The entanglement entropy is defined as the von-Neumann entropy corresponding to the reduced density matrix $\rho$
\begin{align}
S(\rho) = - \tr ( \rho \log \rho  ).
\end{align}

\subsection{The first law}
The definition of the relative entropy can be recast into the following form 
\begin{align}
\S (\sigma||\rho)  = \frac{1}{T} ( F(\sigma)  - F (\rho)).
\end{align}
Here, $F(\rho)$ is the free energy
\begin{align}
F(\rho)  = \tr (\rho H)  - T S(\rho).
\end{align}
The relative entropy can then be expressed as (implicitly absorbing the factor of $1/T$ in $H$)
\begin{align} \label{RE-therm}
\S (\sigma ||\rho ) = \Delta \langle H \rangle- \Delta S,
\end{align}
where, $\Delta \langle H \rangle = \tr (\sigma H) - \tr(\rho H)$  and $\Delta S = S (\sigma)-S(\rho)$.

Using the non-negativity property of the relative entropy ($\S(\sigma || \rho) \geq 0$), we obtain
\begin{align}\label{1-law}
\Delta \langle H \rangle\geq \Delta S .
\end{align}
The above inequality sets a bound on the variation of entanglement entropy between two states. The variation in EE is always lesser than the variation of the modular Hamiltonian. When the inequality of \eqref{1-law} is saturated, it is often referred to as the first law of entanglement \cite{Faulkner:2013ica,Lashkari:2013koa}. 

We shall now investigate two cases for which the inequality \eqref{1-law} gets saturated.
\begin{enumerate}
\item Consider a parameter $\lambda$ which labels the states of the system $\rho(\lambda)$. Let $\lambda=0$ denote the ground state.
The first order variation in $\lambda$ of the entanglement entropy is
\begin{align} \label{pert-1st-law}
\delta S = - \tr (\delta \rho \log \rho )  - \tr (\rho \rho^{-1} \delta \rho) = \tr (\delta\rho H) = \delta \langle H
\rangle .
\end{align} 
Thus, the equality is true to first order in perturbation theory in $\lambda$. 

\item It was noted before, that the relative entropy vanishes in the limit of small sub-system sizes. Thus, combining \eqref{small-size} and \eqref{RE-therm} we obtain for a sub-system $A$
\begin{align} \label{non-pert-1st-law}
\lim_{{\frac{\dim(A)}{\dim{(A')}} \rightarrow 0}} \ \big(\Delta \langle H_A \rangle - \Delta S_A \big) =0 ,
\end{align}
where, $H_A$ and $S_A$ denote the modular Hamiltonian and the entanglement entropy respectively. 
\end{enumerate}

In the holographic context, \eqref{pert-1st-law} was verified in \cite{Blanco:2013joa}  for a number of examples in arbitrary dimensions. It was also shown in \cite{Blanco:2013joa}, that \eqref{non-pert-1st-law} holds good to leading order in sub-system sizes for 1+1 dimensional holographic CFTs. 
\begin{align} \label{blanco}
&\Delta S \Big|_{\text{to }\phi^2} \ \bm{=} \ \Delta \H \Big|_{\text{to }\phi^2} 
= \c \, \frac{ \left((\ell T)^2 + 1\right)}{72}  \, \phi^2 .
\end{align}
This was shown by calculating the differences in $\H$ and $S$ corresponding to the BTZ and global $AdS_3$ geometries -- these backgrounds are dual to a high  temperature state and the vacuum of the dual CFT.  

Equation \eqref{small-size} and \eqref{non-pert-1st-law} can be understood from a holographic viewpoint\footnote{We thank Lin-Yan Hung for discussions regarding this point.}. A state in the CFT Hilbert space corresponds to a specific bulk configuration. However, for  sufficiently small entangling intervals, the Ryu-Takayanagi minimal surface is insensitive about the details of the bulk geometry (i.e.~presence of horizons etc.), since it is very close to the boundary of $AdS$. This forces the entanglement entropies (or equivalently the density matrices) to become close to that of the vacuum state. 

   In this work, we explicitly verify \eqref{non-pert-1st-law} for systems with a non-zero chemical potential $(\mu)$ corresponding to a higher spin current. We show that at leading order in sub-system sizes, $\Delta\langle H_A \rangle = \Delta S_A$. This will involve a generalization of \eqref{blanco} for non-zero $\mu$.

\subsection{Modular Hamiltonians in CFTs}

For spherical entangling surfaces,  $S^{d-2}$, in a $d$-dimensional conformal field theory,  the modular Hamiltonian associated to the CFT vacuum can be expressed in terms of the stress tensor \cite{Myers:2010tj,Casini:2011kv}
\begin{align}
H_{\text{vac}} = 2\pi \int _{|x| < R} d^{d-1} x \, \frac{R^2 - r^2}{2R} T_{00}(\vec{x}) .
\end{align}
One can obtain the modular Hamiltonian for $R\times H^{d-1}$ by conformal transformation of the above formula. This is followed by a coordinate transformation which maps $R\times H^{d-1}$ to the cylindrical geometry $R\times S^{d-1}$. Performing this series of transformations in the $1+1$ dimensional CFT results in the following expression for the modular Hamiltionian 
\begin{align}\label{modular-formula}
H_{\text{vac}}= 2\pi R^2 \int_{-\frac{\phi}{2}}^{\frac{\phi}{2}} \, d\theta \, \frac{\cos \theta - \cos\frac{\phi}{2}}{\sin\frac{\phi}{2}} T_{00}(\theta) ,
\end{align}
where, we have considered a finite size system of length $R$ and the entangling interval is of length $\phi R$. The energy density $T_{00}$ can also be written as $(L_0 - \tfrac{c}{24})+ (\bar{L}_0 - \tfrac{c}{24})$. 

One can use \eqref{modular-formula}, which is an operator relation, to calculate the expectation values of the vacuum state and the high-temperature state
\begin{align*}
\langle H\rangle _{\text{vac}} = \text{Tr}(\rho _{\text{vac}}H_{\text{vac}}) \ , \qquad \langle H\rangle _{\text{T}} = \text{Tr}(\rho _{\text{T}}H_{\text{vac}}) 
\end{align*}
It is important to note that in both expectation values the modular Hamiltonian is that of the vacuum, however they are weighted with the density matrices of the corresponding states. 
\def\vcm{{\text{vac}}}
\vspace{0.2cm}

We shall be interested in calculating the quantities mentioned above for a CFT with $\cW_3$ symmetries at finite chemical potential for the spin-3 current. The relative entropy shall be calculated between a high temperature state and the vacuum. Let us explicitly mention the thermodynamic equality for relative entropy which we shall be interested in.
\begin{align}
\S (\rho_T || \rho_\vcm) = (H_T - H_\vcm) - (S_T - S_\vcm )
\end{align}
Here, $S_T$ and $S_\vcm$ are the entanglement entropies of the single interval of size $R\phi$ corresponding to the high temperature state and the vacuum. 

In the next section we shall examine how these quantities are calculated from the holographic setup.

\section{Entanglement entropy from higher spin holography} \label{hs-ee}
%

In this section we shall review the Chern-Simons formulation of higher spin gravity in 2+1 dimensional spacetime. We then review the Wilson line proposal  of  \cite{deBoer:2013vca} for holographic entanglement entropy. The holographic entanglement entropy is then calculated corresponding to the $SL(3,\mathbb{R})$ higher spin black hole and the higher spin vacuum in the bulk.
 We also evaluate the entanglement entropy corresponding to a smooth solution $\gamma$ in the $SL(2,\mathbb{Z})$ family of solutions. 

\subsection{Basics of 3\textit{d }higher spin gravity}

The action describing Einstein gravity with a negative cosmological constant in three dimensions can written in terms of the difference of two Chern-Simons actions as follows
\begin{align}\label{cs-action}
&I = I_{CS} [A] - I_{CS} [\bar{A}],  \nn \\
&\text{where, }  I_{CS} (A) = \frac{k}{4\pi}  \int_{\Sigma} \tr \left(  A \wedge dA + \frac{2}{3}  A \wedge A \wedge A   \right) .
\end{align}
Here, $A\ (=A_\mu ^a T^a)$ and $\bar{A}$ are $SL(2,\mathbb{R})$ valued gauge connections which are linear combinations of the tetrad and vielbein
\begin{align}
A=\omega + e \quad , \quad \bar{A} = \omega - e \quad , \quad g_{\mu\nu} = \frac{1}{2} \left(  e_\mu e_\nu   \right).
\end{align}
The Chern-Simons level $k$ is related to the radius of $AdS_3$ and the Newton's constant as
\begin{align}
k = \frac{l_{AdS_3}}{4G_N^{(3)}}.
\end{align}
It can also be expressed in terms of central charge of the dual CFT, via the famous Brown-Henneaux relation, $k=\c/6$.

Analogously, if we want describe a gravity theory with higher spins ($s= 2, \cdots, N$), the gauge group  in \eqref{cs-action}  is then chosen to be $SL(N, \mathbb{R})$. This facility of a consistent truncation to a finite tower of spins is only possible in 3\textit{d} gravity. There exist various embeddings of the $SL(2, \mathbb{R})$ in $SL(N, \mathbb{R})$.    We shall be restricting ourselves to the  principal embedding  which will be sufficient for the purposes of this paper. 

Vasiliev theory in three dimensions consists of an infinite tower of spins ($s\geq2$) and a massive scalar. The gauge sector can be consistently decoupled from the scalar sector and is describable by the action in \eqref{cs-action} with the gauge group hs$[\lambda]$. 

\subsection{Classical solutions in $SL(3,\mathbb{R})$ Chern-Simons theory} \label{class-sol}

We shall now review classical solutions in $SL(3,\mathbb{R})\oplus SL(3,\mathbb{R})$ Chern-Simons theory.
We include this for the sake of completeness and the expressions of the quantities which we shall write below shall be relevant for the calculation of entanglement entropy and the modular Hamiltonian corresponding to these backgrounds.

 It is well known that the equations of motion of \eqref{cs-action} read
$$
F(A)=0=F(\bar{A}) \qquad \text{where, } F(A) = dA + A \wedge A. 
$$
Thus, all classical solutions in Chern-Simons theory are flat connections.  We shall describe higher-spin black holes and the higher spin vacuum  (at non-zero spin-3 chemical potential) which are higher-spin generalizations of the BTZ black hole and global $AdS_3$. All classical solutions admit a gauge invariant characterization, in terms the holonomy along specific cycles. 

The generators of  $sl(3,\mathbb{R})$ are denoted by $\lbrace L_{0,\pm 1}  , W_{0,\pm 1,\pm 2}  \rbrace$.

\subsubsection*{Higher spin black holes}
 
Higher spin black holes were constructed in \cite{Gutperle:2011kf}. They can be described by the following gauge connections ($x^\pm= t\pm \phi$ and $\phi \to \phi + 2\pi$)
\begin{align}
A = b^{-1} a(x^+) b +  b^{-1} d b  \quad , \quad \bar{A}= \bar{a} (x^-) b^{-1} +b db^{-1},
\end{align} 
where, $b=e^{\rho L_0}$ and
\begin{align}\label{gauge-connections}
 a=& \left(  L_{1}   - \frac{2\pi\cL}{k}  L_{-1} - \frac{\pi }{2k} \cW W_{-2}    \right) dx^+ 
\nn \\ & \qquad
+ \mu \left( W_2 - \frac{4\pi \cL}{k} W_0 + \frac{4\pi^2 \cL^2}{k^2} W_{-2} +  \frac{4\pi \cW}{k} L_{-1}  \right)dx^-, \\
 \bar{a} =& -\left(  L_{-1}   - \frac{2\pi\bar\cL}{k}  L_{1} - \frac{\pi }{2k} \bar\cW W_{2}    \right) dx^-
\nn \\ & \qquad 
- \bar\mu \left( W_{-2} - \frac{4\pi \bar\cL}{k} W_0 + \frac{4\pi^2 \bar\cL^2}{k^2} W_{2} +  \frac{4\pi \bar\cW}{k} L_{1}  \right)dx^+. \nn
\end{align}
We shall restrict ourselves to the static case which has $\bar\mu=-\mu$, $\bar{\cL}=\cL$ and $\bar{\cW}=-\cW$. 

The gauge invariant characterization of the black hole solutions above is given in terms of the holonomies of connections. In the Euclidean signature ($x^+ \to z=\phi+i t_E$ and $x^- \to \bz=\phi-i t_E$), black holes have their thermal/time cycle is contractable in the bulk (the base space being that of a solid torus). The holonomy along the thermal cycle $(z,\bar{z}) \rightarrow (z+2\pi \tau , \bar{z}+2\pi \bar \tau )$ is 
\begin{align}
\text{Hol}_\tau [A]  = \mathcal{P} \exp \left( \oint_\tau A  \right)   =b^{-1} e^{\omega} b \quad , \quad \omega_\tau = 2\pi (\tau a_z - \bar{\tau} a_{\bar{z}}).
\end{align}
This holonomy is then imposed to be the same as that of the BTZ black hole principally embedded in $sl(3,\mathbb{R})$, i.e.
\begin{align}
\text{spec}(\omega_\tau)= \text{spec}((\omega_\tau)_{BTZ}) = (0,\pm 2 \pi i).
\end{align}
The following conditions can be imposed as 
\begin{align}
\det (\omega_\tau ) =0 \quad , \quad \tr (\omega_\tau^2) =- 8 \pi^2  .
\end{align}
There exist four solutions to the equations above which correspond to different thermodynamical branches of the higher spin black hole \cite{David:2012iu,Chen:2012ba,Chowdhury:2013roa}. We shall confine our attention to the branch that is smoothly connected to the BTZ black hole as $\mu \rightarrow 0$. The charges $\cL$ and $\cW$ for this solution can be expressed perturbatively in $(\pi\mu T)$ 
\begin{align} \label{hsbh-l}
\cL =& \ \frac{ \c \pi   T^2 }{12} \Bigg[ 1+\frac{80(\pi   \mu  T)^2}{3}  +\frac{2560 (\pi   \mu  T)^4 }{3}  +\frac{905216(\pi   \mu  T)^6}{27} \\ \nn & \qquad \qquad +\frac{118095872(\pi   \mu  T)^8}{81} + \frac{5475663872(\pi   \mu  T)^{10}}{81}  + \cdots \Bigg], \\
\cW =& \ \frac{8 \c \pi ^2  T^3}{9} 
\Bigg[     \pi  \mu  T + \frac{320(\pi   \mu  T)^3}{9} +\frac{4352(\pi   \mu  T)^5}{3} +\frac{1736704(\pi   \mu  T)^7}{27}    \\ \nn & \qquad \qquad \qquad
 + \frac{244449280(\pi   \mu  T)^9}{81}  + \cdots \Bigg].
\end{align}
The thermodynamics of the such black holes have been extensively studied in a number of works (see \cite{Perez:2014pya} and references therein). It's worthwhile mentioning at this point that chemical potentials conjugate to the higher spin charges $\cW$ and $\bar{\cW}$ are $\alpha$ and $\bar{\alpha}$ given by 
\begin{align}\label{chem-pot-def}
\alpha = \bar{\tau}  \mu  \quad , \quad \bar{\alpha}= \tau \bar{\mu}
\end{align}
where $\tau=(i\beta+i \beta \Omega)/(2\pi)$. For the non-rotating case $\Omega=0$ and $\alpha =\mu/(2\pi i T)$. 

It has been confirmed by CFT computations \cite{Gaberdiel:2012yb,Kraus:2011ds}, that such black holes when appropriately generalized to hs$[\lambda]$, describe a $\cW_\infty[\lambda]$ CFT at high temperatures in the presence of a spin-3 chemical potential\footnote{Black hole solutions in higher spin gravity  with refined boundary conditions and definitions of charges was recently constructed in \cite{Bunster:2014mua}. It shall be interesting to perform the calculations of this paper in those backgrounds.}. 

%

\subsubsection*{Higher spin vacuum}

The higher spin vacuum is a generalization of the global-$AdS_3$ solution for non-vanishing spin-3 chemical potential -- but still at zero-temperature. This solution was constructed in \cite{Compere:2013nba}. As the name implies, solution serves as a holographic description of the vacuum in a $\cW_3$ CFT which is sensitive a spin-3 chemical potential.

The flat connection for this solution is the same as that of  \eqref{gauge-connections}. However, the crucial difference in this case is that the holonomy condition is imposed along the angular ($\phi$) direction  
\begin{align}
\text{Hol}_\phi [A] = \mathcal{P} \exp \left( \oint_\phi A  \right)   =b^{-1} e^{\omega} b \quad , \quad \omega_\tau = 2\pi ( a_z -   a_{\bar{z}}).
\end{align}
This holonomy is imposed to be the same as that of global $AdS_3$ which has a contractable $\phi$-cycle (the boundary being that of a cylinder, with a periodic spatial direction) 
\begin{align}
\text{spec}(\omega_\phi)= \text{spec}((\omega_\phi)_{AdS_3}) = (0,\pm 2 \pi i).
\end{align}
Once again, one imposes the above constraint in a similar manner
\begin{align}
\det (\omega_\phi ) =0 \quad , \quad \tr (\omega_\phi^2) =- 8 \pi^2  
\end{align}
The above equations admit two real solutions \cite{Compere:2013nba,Chowdhury:2013roa}. Just like the case of higher spin black holes we restrict our attention to the branch which is smoothly connected to the one in the $SL(2,\mathbb{R})$ theory i.e.~$AdS_3$.
 The charges then take the following form\footnote{Here $2\pi R$ is the length of the spatial cycle. One can think of the CFT being on a finite size system.}
 \begin{align}\label{hscs-l}
 \cL_\vcm=\ & -\frac{\c}{48 \pi  R^2} \Bigg[   1 -\frac{20 }{3 } \left(\frac{\mu}{R}\right)^2+\frac{160 }{3  }\left(\frac{\mu}{R}\right)^4-\frac{14144  }{27  }\left(\frac{\mu}{R}\right)^6+\frac{461312 }{81 }\left(\frac{\mu}{R}\right)^8  \nn \\ & \qquad\qquad \qquad -\frac{5347328 }{81 }\left(\frac{\mu}{R}\right)^{10} + \cdots \Bigg] , \\
  \cW_\vcm=\ & \frac{\c}{18 \pi  R^3} \Bigg[  \left(\frac{\mu}{R}\right)    -\frac{80  }{9 }  \left(\frac{\mu}{R}\right)^3  +\frac{272  }{3  }  \left(\frac{\mu}{R}\right)^5   -\frac{27136 }{27  }  \left(\frac{\mu}{R}\right)^7  + \frac{954880  }{81 }  \left(\frac{\mu}{R}\right)^9+ \cdots  \Bigg]. \label{hscs-w}
 \end{align}
It was shown in \cite{Compere:2013nba} that the number of gauge symmetries preserved for this branch is the same as that of the global $AdS_3$ and is therefore maximally symmetric. One can therefore expect this bulk configuration to describe the vacuum of a large-$\c$ CFT with a $\cW_3$ symmetry at non-vanishing spin-3 chemical potential. 



It is important to emphasize at this point that one could have also considered solutions in the Euclidean theory $SL(3,\mathbb{C})$ -- these are referred to as conical surpluses. In such a case the temporal direction is made periodic but is still non-contractable. This is a higher-spin generalization of the thermal $AdS_3$ and would describe the dual CFT at low temperatures (below the Hawking-Page phase transition \cite{Chowdhury:2013roa}).
 However, it is crucial to note that even at these low temperatures the holonomy condition is still the same as that of the higher spin vacuum and the expressions for the charges \eqref{hscs-l} and \eqref{hscs-w} are still valid in the large-$\c$ regime. 

\subsubsection*{The $SL(2,\mathbb{Z})$ family of smooth solutions}

It was shown in \cite{Li:2013rsa} that the Euclidean conical surplus and the black hole we just described in the preceding sub-sections are special cases of a more generic family of solutions characterized by the modular parameter of the boundary torus, $\hat{\gamma}\tau$
\begin{align}
 \hat{\gamma}\tau \equiv \frac{a\tau + b}{c \tau + d}\ , \quad \text{with } \gamma=\begin{pmatrix}
 a &b \\
 c &d
 \end{pmatrix} \in PSL(2, \mathbb{Z}) .
 \end{align}
The geometry of the base space of these solutions is the solid torus. The contractible ($\A$) and non-contractible ($\B$) cycles of the torus are 
\begin{align}
\A\text{-cycle } & : z \sim z + (c \tau + d), \nn \\
\B\text{-cycle } & : z \sim z + (a \tau + b) .
\end{align}
The holonomies around these cycles are ($ \text{Hol}_C [A]= b^{-1} e^{2 \pi \omega_C} b $ and $\omega_C = \frac{1}{2\pi} \oint_C a$)
\begin{align}
\omega_{\A} = (c \tau + d) a_z + ( c\bar{\tau} +d   )a_{\bar{z}}, \nn \\
\omega_{\B}= (a \tau + b) a_z + (a\bar{\tau} +b )a_{\bar{z}} .
\end{align}
The holonomy condition can then be imposed such that it belongs to center of the group $SL(N,\mathbb{R})$. We shall be imposing the condition
\begin{align}
\text{spec}(\omega_{\A}) = 2\pi i  \vec{\rho} = (-2\pi i , 0 , 2 \pi i) .
\end{align}
 where $\vec{\rho}$ is the Weyl vector of $SL(3,\mathbb{R})$ which corresponds the holonomy of global $AdS_3$ along the $\phi$-cycle. 
 
 It can now be easily seen that the Euclidean conical surplus is a special case of the above with $a=1,\ b=0,\ c=0,\ d=1$ while the black hole has $a=0,\ b=-1,\ c=1,\ d=0$.
 
%
%
 The spin-2 and spin-3 charges can we written as 
 \begin{align} \label{sl2z-l}
 \cL = &-\frac{\c}{48 \pi  (c \tau +d)^2} \Bigg[   1 -\frac{20 }{3} \left( \frac{\mu}{c \tau + d}  \right)^2 +\frac{160  }{3 } \left( \frac{\mu}{c \tau + d}  \right)^4-\frac{14144 }{27}\left( \frac{\mu}{c \tau + d}  \right)^6 \nn \\ & \hspace{4cm}+\frac{461312 }{81 } \left( \frac{\mu}{c \tau + d}  \right)^8-\frac{5347328 }{81} \left( \frac{\mu}{c \tau + d}  \right)^{10}+ \cdots \Bigg] , \\
  \cW = & \frac{\c \, \mu }{18 \pi  (c\tau + d)^4} \Bigg[   1 -\frac{80  }{9 }\left( \frac{\mu}{c \tau + d}  \right)^2+\frac{272  }{3  }\left( \frac{\mu}{c \tau + d}  \right)^4-\frac{27136 }{27  }\left( \frac{\mu}{c \tau + d}  \right)^6\nn \\ & \hspace{4cm} +  \frac{954880 }{81 }\left( \frac{\mu}{c \tau + d}  \right)^8 + \cdots \Bigg] .
 \end{align}

 \subsection{Holographic entanglement entropy}
 
In a higher spin theory of gravity the metric is not gauge-invariant. Therefore, usual notions of horizons or singularities or minimal surfaces do not form good quantities to describe quantities that are relevant to describe the dual CFT. 
The generalization of the Ryu-Takayanagi minimal area prescription \cite{Ryu:2006bv,Nishioka:2009un} to the case of higher spin gravity was recently provided in  \cite{deBoer:2013vca} and \cite{Ammon:2013hba}. The appropriate gauge-invariant observable (akin to geodesics for the case of $AdS_3$) is that of the Wilson line.

Since our calculation of the charges is in the holomorphic formalism, we shall use the holomorphically factorized version of Wilson line functional of \cite{deBoer:2013vca}. The holomorphic prescription has also been verified perturbatively by CFT computations in \cite{Datta:2014ska,Datta:2014uxa}. 

According to \cite{deBoer:2013vca}, the holographic entanglement entropy for the subsystem defined by the open-interval $(P,Q)$ living on the   boundary of $AdS_3$ is given by the functional
\begin{align}
S_{(P,Q) }= \frac{\c}{\sigma_{\mathcal{R}}} \log \left[   \lim_{\rho_0 \rightarrow \infty  } W_{\mathcal{R}} (P,Q) \Big|_{\rho_P=\rho_Q=\rho_0}   \right],
\end{align}
where,
\begin{align}
W_{\mathcal{R}} (P,Q) \equiv \tr_{\mathcal{R}}  \Bigg[ \mathcal{P} \exp \left(\int_{P}^{Q} \bar{A}_{-} dx^- \right)  \mathcal{P}  \exp \left(\int_{Q}^{P} {A}_{{+}}dx^+  \right) \Bigg].
\end{align}
The representation is chosen such the entanglement entropy reduces to the thermal entropy in the extensive limit $\Delta x  \gg \beta$. For the specific case of $SL(3,\mathbb{R})$ gravity, $\mathcal{R}$ is the 8-dimensional adjoint representation and $\sigma_{\mathcal{R}}=24$. 
 \def\F{\mathcal{F}}
For gauge-connections of the form \eqref{gauge-connections} it can then be shown that the Wilson line takes the form (generally for the non-static case),
\begin{align}\label{hee-formula}
  \lim_{\rho_0 \rightarrow \infty  } W_{\mathcal{R}} (P,Q) \Big|_{\rho_P=\rho_Q=\rho_0}  = \F (\lambda_1,\lambda_2,\lambda_3)  \F (\bar \lambda_1,\bar \lambda_2,\bar \lambda_3) ,
\end{align}
 where,
 \begin{align}
  \F (\lambda_1,\lambda_2,\lambda_3) =   \frac{8 \Lambda^4}{{{ \lambda_1} { \lambda_2} { \lambda_3}}} { \left(\frac{{ \lambda_1}^2-{ \lambda_2} { \lambda_3}}{{ \lambda_1} { \lambda_2} { \lambda_3}} +\frac{\cosh (\Delta  { \lambda_1})}{{ \lambda_1}}-\frac{\cosh (\Delta  { \lambda_2})}{{ \lambda_2}}-\frac{\cosh (\Delta  { \lambda_3})}{{ \lambda_3}}\right)} .
 \end{align}
  Here $\lambda_i$ and $\bar{\lambda}_i$ are the eigenvalues of $a_z$ and $\bar{a}_\bz$ in the adjoint representation. $\Delta$ is the length of the interval (distance between $P$ and $Q$, $|z_{PQ}|$). $\Lambda^{-1}(=e^{-\rho_0})$ is the UV cutoff.  We shall be using \eqref{hee-formula} to calculate the entanglement entropy corresponding the higher spin vacuum, the higher spin black hole and  smooth solutions in the $SL(2,\mathbb{Z})$ family. 
  
  A suitable parametrization of the charges and chemical potentials was introduced in \cite{Ammon:2011nk}
  \begin{align}
  \cW = \frac{4(C-1)}{C^{3/2}} \cL \sqrt{\frac{2\pi \cL}{k}} \ , \quad \mu = \frac{3\sqrt{C}}{4(2C-3)} \sqrt{\frac{k}{2\pi \cL}} \ , \quad \tau = \frac{i (2C-3)}{4(C-3)\sqrt{1- \frac{3}{4C}}} \sqrt{\frac{k}{2\pi \cL}} \nn .
  \end{align}
The eigenvalues in the adjoint representation ($0,\pm \lambda_1\pm \lambda_2,\pm \lambda_3$) then take the following form 
\begin{align} \label{eigenvalues}
&\lambda_1 = 4 \sqrt{\frac{2\pi \cL}{k}} \sqrt{1- \frac{3}{4C}} \ , \ \nn \\ &\lambda_2 =  2 \sqrt{\frac{2\pi \cL}{k}} \left(\sqrt{1- \frac{3}{4C}}-\frac{3}{2\sqrt{C}} \right) \ , \ \lambda_3 =  2 \sqrt{\frac{2\pi \cL}{k}} \left(\sqrt{1- \frac{3}{4C}}+\frac{3}{2\sqrt{C}} \right) .
\end{align}
The parameter $C$ can in-turn  be solved in terms of $\cL$ as (it has got only one real solution)
\begin{align}\label{see}
C=\frac{3}{512\c \pi \mu^2 \cL} \left( \c + 256 \pi \mu^2 \cL + \sqrt{\c^2+512\pi \c \mu^2 \cL }   \right) .
\end{align}
Until this point in the calculation, no approximations have been made and the treatment is general for the higher spin vacuum and  for all solutions in the $SL(2,\mathbb{Z})$ family without any reference to holonomy conditions. 

\subsubsection*{EE corresponding to higher spin black holes}
We shall now find the entanglement entropy in the regime of small chemical potentials i.e.~perturbatively in $\mu$. Substituting the solutions to holonomy conditions \eqref{hsbh-l} in \eqref{see} and then evaluating the eigenvalues and finally substituting them in the Wilson line formula \eqref{hee-formula}, we obtain
\begin{align}\label{ee-bh}
S_T (\phi)= & \frac{\c}{3}\log \Big| \frac{\sinh (\pi  R T \phi )}{ \Lambda^{-1} \, \pi T }\Big| \nn \\ &+ \frac{\c}{18} (\pi\mu  T)^2 \text{csch}^4(\pi  R T \phi ) \left. \Big[ 8\left(1-3 \pi ^2 R^2 T^2 \phi ^2\right) \cosh (2 \pi  R T \phi )\right.\nn \\ &\hspace{5.5cm} \left.+8 \pi  R T  \phi\left(  \sinh (2 \pi  R T \phi )+  \sinh (4 \pi  R T \phi ) \right) \right.\nn \\ & \hspace{5.5cm } \left.-5 \cosh (4 \pi  R T \phi )-3\right. \Big] + \order((\pi \mu T)^4) .
\end{align}
One can systematically keep track of terms to any arbitrary order in  $(\pi \mu T)$. Although we shall be working to higher orders, we do not quote the expressions here as they are large and are not particularly illuminating. 

The above formula holographically describes a $\cW_3$ CFT at large central charge $\c$  deformed by a chemical potential for the spin-3 current at high temperatures\footnote{The minimal-model/higher spin dualities relate $\cW_\infty[\lambda]$ CFTs  (for $0\leq\lambda\leq 1$) to higher spin gravity hs$[\lambda]$ in the bulk along with two scalars with specific masses.  The exact CFT dual for $SL(N,\mathbb{R})$ gravity is unknown. However, the asymptotic symmetries of $SL(3,\mathbb{R})$ gravity is that of the $\cW_3$-algebra at large central charge and also $SL(N,\mathbb{R})$-black hole partition functions match with the high temperature $\cW_N$ CFT partition function for $\lambda = \pm N$. Moreover, hs$[\lambda]$ reduces to $SL(N,\mathbb{R})$ for $\lambda=-N$ after factoring out the ideal. For these reasons, we shall be abstractly referring to the boundary theory as `a CFT with a $\cW_3$ symmetry at large central charge'.   See \cite{Perlmutter:2012ds} for a version of the duality at finite $N$, although the CFT is non-unitary.}. 

\subsubsection*{EE corresponding to higher spin vacuum}
In a similar manner, the entanglement entropy corresponding to the higher spin vacuum solution can also be evaluated. The expression for the EE to order $(\mu /R)^2$ is 
\begin{align}\label{ee-cs}
S_{\text{vac}} (\phi) = & \frac{\c}{3}\log \Big| \frac{2 R}{\Lambda^{-1}} \  {\sin\left(\frac{\phi}{2} \right)}   \Big| \nn \\
 & +\frac{\c}{72} \left(\frac{\mu}{R} \right)^2 \csc ^4\left(\frac{\phi }{2}\right)\Big[ 3 -2 \left(3 \phi ^2+4\right) \cos (\phi )+4 \phi ( \sin (\phi )+  \sin (2 \phi ) )  \nn \\ & \hspace{4.4cm}+5 \cos (2 \phi ) \Big] + \order((\mu/R)^4) .
\end{align}
Needless to say, one can also retain terms to an arbitrary order in $(\mu/R)$. 
This expression is strictly true when the dual CFT is in the  vacuum state or at $T=0$. As emphasised earlier for the case of the charges $\cL_\vcm$ and $\cW_\vcm$, the entanglement entropy in \eqref{ee-cs} is even valid for low temperatures (lesser than the Hawking-Page temperature). 

It may be noted here, that even if the CFT is at $T=0$, the property $S_A= S_{A'}$ or $S(\phi)=S(2\pi - \phi)$ is not obeyed for finite chemical potential. This is due to the presence of non-periodic functions in the second term of \eqref{ee-cs}. This shows the state doesn't remain pure when the chemical potential is switched on. 

\def\sl2{{SL(2,\mathbb{Z})}}

\subsubsection*{EE corresponding to smooth solutions in the $SL(2,\mathbb{Z})$ family}
 The holographic entanglement entropy corresponding a member $\gamma$ in the $\sl2$ family of smooth solutions can also also be calculated. We use the perturbative expansion in $\mu$ for $\cL$ given in \eqref{sl2z-l}. We use $\xi \equiv c\tau + d$ for brevity. 
\begin{align}\label{ee-sl2}
S_{\hat\gamma} (\phi) = &  \frac{\c}{3}\log \Big| \frac{2 \xi}{\Lambda^{-1}} \  {\sin\left(\frac{R \phi}{\xi} \right)}   \Big| \nn \\
& + \frac{\c}{72} \left(\frac{\mu}{\xi}\right)^2 \csc ^4 \left( \frac{R \phi}{2\xi } \right) \Bigg[ 3-2\left( 4 + 3  \left(\frac{R \phi }{\xi }\right)^2 \right) \cos \left(\frac{R \phi }{\xi }\right) +5 \cos \left(\frac{2 R \phi }{\xi }\right)  \nn \\
& \hspace{4.6cm}+4  \left(\frac{R \phi }{\xi }\right) \left(  \sin \left(\frac{R \phi }{\xi }\right)+\sin \left(\frac{2 R \phi }{\xi }\right)  \right) 
 \Bigg] \nn \\ & \qquad \hspace{8.5cm}+ \order ((\mu/\xi)^4). 
\end{align}
The above expression reduces to that of the conical surpluses for $c=R$ and $d=0$ and that of the black hole for $c=0$ and $d=1$. 

\section{Relative entropy of holographic CFTs with a $\cW_3$ symmetry}
\label{rel-calc}
We shall now utilize the results of the previous section to find the holographic relative entropy between  a high-temperature state  and the vacuum in a $\cW_3$ CFT  deformed by a spin-3 chemical potential. We shall also check the validity of the first law of entanglement, $\Delta H = \Delta S$, in the regime of small sub-system sizes \eqref{non-pert-1st-law}.  

Since we have also evaluated the expressions for the holographic entanglement entropy for a smooth solution in the $\sl2$ family, we can find the relative entropy between such a solution $\gamma$ and the higher spin vacuum.  
\def\vac{{\text{vac}}}
\subsection{Relative entropy between the vacuum and high temperature states}

In \cite{Blanco:2013joa} the relative entropy was evaluated between a high temperature state and the vacuum of  $(1+1)d$ CFTs. It was also shown that the relation $\Delta H = \Delta S$ is true to the leading order of sub-system sizes. We shall now try to generalize those results in the simplest theory of higher-spin gravity -- the $SL(3,\mathbb{R})$ Chern-Simons theory. 

We shall employ the thermodynamic-like relation \eqref{RE-therm} to calculate $\S (\rho_{T}||\rho_{\text{vac}})$
\begin{align}
\S (\rho_{T}||\rho_{\text{vac}}) = (\langle H_T \rangle - \langle H_\vac \rangle) - (  S_T - S_\vac  ).
\end{align}
All quantities above are evaluated for the subsystem of size $R\phi$. 

\subsection*{Change in entanglement entropy $\Delta S$}
 As mentioned previously, a high temperature state and the vacuum of the $\cW_3$ CFT admit holographically dual configurations in terms of the higher spin black hole and the higher spin vacuum respectively. The difference in entanglement entropies of the corresponding states is 
\begin{align}
\Delta S (\phi) = S_T (\phi) - S_{\text{vac}} (\phi).
\end{align}
Here we shall directly  use the expressions for $S_T (\phi)$ in \eqref{ee-bh} and $S_{\text{vac}} (\phi)$ in \eqref{ee-cs}. Expanding the result for $\Delta S (\phi) $ for short sub-system/interval sizes ($\phi \ll 1$)  (and keeping track of terms to some higher orders in $\mu/R$) we obtain the following
\begin{align} \label{delta-S}  \nn
\Delta S (\phi)=& \ \c \Bigg[ \frac{ \left((2\pi R T)^2 + 1\right)}{72} 
+
\frac{5\left((2\pi R T)^4-1\right)}{54}  \left(\frac{\mu}{ R}\right)^2
+
\frac{20 \left((2\pi R T)^6+1\right)}{27}  \left(\frac{\mu}{ R}\right)^4
\\ \nn & \qquad+
\frac{1768  \left((2\pi R T)^8-1\right)}{243}  \left(\frac{\mu}{ R}\right)^6
+
\frac{57664 \left((2\pi R T)^{10}+1\right)}{729 } \left(\frac{\mu}{ R}\right)^8
 \\ 
& \qquad + \frac{668416  \left((2\pi R T)^{12}-1\right)}{729 }  \left(\frac{\mu}{ R}\right)^{10} +  \mathcal{O}\left( \left(\mu/R\right)^{12}\right)   \Bigg] (\phi)^2 \ + \ \mathcal{O}((\phi) ^4).
\end{align}
It is obvious that for $\mu=0$, the expression reduces to that of $\Delta S (\phi)$ evaluated between the BTZ and global $AdS_3$ geometries \cite{Blanco:2013joa}.  

\subsection*{Change in the modular Hamiltonian $\Delta H$}

We shall now calculate the difference in the expectation values of the modular Hamiltonian $\Delta \H$. We shall make use of the formula \eqref{modular-formula} 
\begin{align} \label{mod-formula-2}
\langle H\rangle _{\rm{state}} = 2\pi R^2 \int_{-\frac{\phi}{2}}^{\frac{\phi}{2}} \, d\theta \, \frac{\cos \theta - \cos\frac{\phi}{2}}{\sin\frac{\phi}{2}} T^{\rm{state}}_{00}(\theta) .
\end{align}
Note that this formula is true for a conformal field theory. 

It was shown in \cite{Compere:2013gja}, that when a spin-3 chemical potential is turned on for the black hole in the principal embedding of $sl(3,\mathbb{R})$,  the asymptotic symmetry algebra is still that of $\cW_3 \times \cW_3$. The black hole was considered with Dirichlet boundary conditions which specified the fall-off conditions at $\rho \to \infty$ and initial data for the higher spin charges ($\cL(0,\phi)$ and $\cW(0,\phi)$ on the Cauchy surface at $t=0$). When a chemical potential is perturbatively turned on, it can be shown by a proper rewriting of the generators that the $\cW_3\times\cW_3$ asymptotic symmetry is still preserved
\footnote{It was shown in \cite{Compere:2013nba} that the canonical definitions of the zero modes in \cite{Compere:2013gja} are related to the charges in the holomorphic formalism through the so-called `tilded variables'. Also, the thermodynamic quantities in the canonical formalism match with those of the holomorphic formalism (and that of the CFT) when expressed in terms of these tilded variables. 
Additionally, \cite{Compere:2013nba} had also shown that the perturbative expansion in $\mu$ has a finite radius of convergence by illustrating the existence of a higher spin vacuum (with exactly the same number of isometries as that of global $AdS_3$) for ${\mu} < \pm \frac{3}{8}  \sqrt{2\sqrt{3}+3} $. 
 }. Since the proof of \cite{Compere:2013gja} did not rely on any reference to holonomy conditions, the asymptotic algebra analysis is quite general and applies to that of the higher spin vacuum solution as well. We shall choose to work with the same boundary conditions in this paper. Furthermore, as mentioned earlier we are focussing on branches that are smoothly connected to the BTZ and global $AdS_3$ where this perturbative analysis of asymptotic symmetries is valid. Since the extended conformal symmetry is unbroken for these branches, it is justified to use \eqref{modular-formula} or \eqref{mod-formula-2} to calculate the modular Hamiltonian even when these higher spin chemical potentials are present. Thus, we are using \eqref{mod-formula-2} which is now generalized to the case of non-vanishing spin-3 chemical potentials. From the CFT perspective, we can evaluate \eqref{mod-formula-2} for states which are held at a finite energy and a finite spin-3 charge by a temperature and a chemical potential. 

The stress tensor component, $T_{00}$, can be  written as follows
\begin{align}
T_{00} = (T_{zz} + T_{\bz \bz})_{\text{CFT}} = (\cL + \bar{\cL})_{\text{holographic}}.
\end{align}
In the first equality we have used the tracelessness condition $T_{z\bz}=0$.
In the second equality we have exploited the AdS/CFT dictionary that the holographic stress tensor, which appears in the Fefferman-Graham expansion of the metric, is equal to that of the CFT \cite{Balasubramanian:1999re,deHaro:2000xn}. (This is true even in the case of higher spin holography \cite{Ammon:2011nk}. Moreover, this way of reading out $\cL$ is independent of whether we choose to work with the canonical or holomorphic formalism.)

For the cases we shall be dealing with, the stress tensors \eqref{hsbh-l}, \eqref{hscs-l} and \eqref{sl2z-l}  do not have any coordinate dependence. We shall also confine our attention to the static case $\bar{\cL}=\cL$ for simplicity. Equation \eqref{mod-formula-2} then simplifies to 
\begin{align}\label{mod-formula-3}
\H_{\text{state}} = 8 \pi  R^2 \left[1-  \frac{\phi }{2} \cot \left(\frac{\phi }{2}\right)\right] \cL_{\text{state}} .
\end{align}
The difference in the modular Hamiltonian between the high-temperature phase (dual to the higher spin black hole) and the vacuum (dual to the higher spin vacuum) is then given by
\begin{align}
\Delta \H = 8 \pi  R^2 \left[1-  \frac{\phi }{2} \cot \left(\frac{\phi }{2}\right)\right] (\cL_{T} - \cL_{\vcm}),
\end{align}
where, we shall be  using the expressions of \eqref{hsbh-l} and \eqref{hscs-l} for $\cL_{T}$ and $\cL_{\vcm}$ respectively. Investigating the behaviour at small entangling intervals, $\phi\ll 1$, we get
\begin{align} \label{delta-H}  \nn
\Delta \H (\phi)=& \ \c \Bigg[ \frac{ \left((2\pi R T)^2 + 1\right)}{72} 
+
\frac{5\left((2\pi R T)^4-1\right)}{54}  \left(\frac{\mu}{ R}\right)^2
+
\frac{20 \left((2\pi R T)^6+1\right)}{27}  \left(\frac{\mu}{ R}\right)^4
\\ \nn & \qquad+
\frac{1768  \left((2\pi R T)^8-1\right)}{243}  \left(\frac{\mu}{ R}\right)^6
+
\frac{57664 \left((2\pi R T)^{10}+1\right)}{729 } \left(\frac{\mu}{ R}\right)^8
 \\ 
& \qquad + \frac{668416  \left((2\pi R T)^{12}-1\right)}{729 }  \left(\frac{\mu}{ R}\right)^{10} +  \mathcal{O}\left( \left(\mu/R\right)^{12}\right)   \Bigg] (\phi)^2 \ + \ \mathcal{O}((\phi) ^4) .
\end{align}
All terms in this perturbative expansion in $\mu/R$, at the leading order  in $\phi$, match exactly with that of $\Delta S(\phi)$ in  \eqref{delta-S}. One can test the matching of the coefficients in \eqref{delta-S} and \eqref{delta-H} to an arbitrary order in $(\mu/R)$. The relative entropy is
\footnote{The detailed expression of the relative entropy at the leading order is 
\begin{align}
\S(\rho_T|| \rho_\vac)\ = \ &\c \Bigg[ \frac{ \left(4 \pi ^2 R^2 T^2+1\right)^2}{8640}+\frac{2}{81} \pi ^4  R^4 T^4  \left(4 \pi ^2 R^2 T^2+1\right) \left(\frac{\mu}{ R}\right)^2\nn \\ &\qquad \qquad +\frac{1}{972} \left(4352 \pi ^8 R^8 T^8+768 \pi ^6 R^6 T^6-5\right)\left(\frac{\mu}{ R}\right)^4   + \cdots  \Bigg]\phi^4 +\order(\phi^6)  \nn .
\end{align}
}
\begin{align}
\S(\rho_{T}||\rho_\vac) = \order(\phi^4).
\end{align}

We have thus established the following via holographic calculations : \textsl{At the leading order in entangling interval sizes, $\Delta H = \Delta S$ in a large-$\c$ CFT with a $\cW_3$ symmetry at finite higher spin chemical potential.} This also verifies that the holographic entanglement entropy, calculated using the Wilson line functional, obeys the equation \eqref{non-pert-1st-law} which is generically  true for a quantum mechanical system.

\subsection*{Discussion}
A few comments are in order, regarding the calculation above and the agreement of $\Delta S$ and $\Delta H$.
\begin{enumerate}

\item It is  interesting to note, that since our CFT is at a non-zero chemical potential, one might expect the na\"{i}ve generalization of the first law of the thermodynamics to entanglement entropy in the regime of small interval sizes
\begin{align}
\Delta S = \Delta \H +  \mu \Delta\langle W\rangle \nn .
\end{align}
Here, $\langle W\rangle$ is the expectation value of some higher spin `modular charge' possibly calculated from the spin-3 charges $\cW$. However, we have seen by explicit computations that this   expectation is not true\footnote{The calculations of relative entropy for $U(1)$-charged black branes in arbitrary dimensions  also lends support in favour of $\delta H = \delta S$  without any corrections \cite{Blanco:2013joa}. The equality was shown for the relative entropy between two neighbouring states in the dual CFT Hilbert space at finite chemical potential.}.  Therefore, the modular Hamiltonian defined via \eqref{modular-formula} encodes the full information of the density matrix $\rho$. Here, $\rho$ may have been defined \textsl{{a} priori} in the grand canonical ensemble.  

\item In the light of recent agreement of the holomorphic Wilson line proposal of \cite{deBoer:2013vca} with CFT computations \cite{Datta:2014ska,Datta:2014uxa}, we have chosen to work in the holomorphic formalism of higher spin gravity. Furthermore, the entanglement entropy to $\order({\mu^2})$ is universal for a $\cW_\infty[\lambda]$ CFT. We therefore have the following (at leading order in small interval size $\phi$)
\begin{align}
\Delta \H (\phi) \Big|_{\text{to }\order({\mu^2})}=& \ \c \Bigg[ \frac{ \left((2\pi R T)^2 + 1\right)}{72} 
+
\frac{5\left((2\pi R T)^4-1\right)}{54}  \left(\frac{\mu}{ R}\right)^2 \Bigg] \phi^2= \Delta S (\phi) \Big|_{\text{to }\order({\mu^2})}
\end{align}
to be true for any $\cW_{\infty}[\lambda]$ CFT. 

\item Note that there are other branches of the higher spin black hole and the higher spin vacuum \cite{David:2012iu,Compere:2013nba,Chowdhury:2013roa,michael}. We have just focussed on the branches which are smoothly connected to the classical solutions in Einstein gravity. The branch of the higher spin black hole reduces to that of the BTZ black hole as $\mu T \rightarrow 0$. For the higher spin vacuum, the branch reduces to that of global-$AdS_3$ as $(\mu/R)\rightarrow 0$. Also, there is an upper bound beyond which these branches cease to exist\footnote{Note that both the upper bounds  automatically restrict the expansion parameters to be less than unity.}.
\begin{align}
\text{BTZ-branch of the higher spin black hole  : }& \pi\mu T < \frac{3}{16}  \sqrt{2\sqrt{3}-3} \ . \nn \\
\text{AdS$_3$-branch of the higher spin vacuum  : }& \frac{\mu}{R}  < \frac{3}{8}  \sqrt{2\sqrt{3}+3} \ . \nn 
\end{align}
This being the case, one cannot verify $\Delta \H = \Delta S$ for arbitrary values of $\mu$ using the above procedure.
Moreover, since these are `the perturbative branches' the asymptotic symmetry algebras of $\cW_3\times \cW_3$ is unbroken as shown in \cite{Compere:2013gja}. It will be however interesting to evaluate the Wilson line functionals corresponding all the branches and thereby investigate the role of entanglement entropy in the phase structure of higher spin gravity. 

\item The calculation of the holographic quantities (namely the stress tensors and entanglement entropies)  have been performed in perturbation theory in $\mu \ll \beta$ or $\mu\ll R$. So, the regime of validity of the expressions \eqref{delta-S} and \eqref{delta-H} is when  $\mu$ is sufficiently small compared the length and temperature scales associated with the CFT. 

\item The verification of $\Delta\H = \Delta S$ also probes the short-distance behaviour of the holographic entanglement entropy. In \cite{deBoer:2013vca}, it had been asserted that one might require to redefine the UV cut-off  in the Wilson line functional at small distances. However, the relative entropy (or rather $\Delta S$) is insensitive to  the UV cut-off and is free of divergences in 2$d$ CFTs. We have explicitly seen that $\Delta S$ evaluated between a  high-temperature state and the vacuum  has the desired behaviour at short distances. 

\item For interval sizes not necessarily short, we expect the inequality $\Delta \H \geq \Delta S$. As remarked earlier, $\order(\mu^2)$ correction to the relative   entropy is universal for any $\cW$-algebra CFT. Since $\cW_\infty[\lambda]$ for $0\leq \lambda \leq 1$ is a unitary CFT, $\Delta \H \geq \Delta S$ should hold true at $\order(\mu^2)$. This can indeed be confirmed numerically. 


\end{enumerate}

\subsubsection*{Considering the global-\textit{AdS} as the vacuum}

In the calculation above, the bulk configuration corresponding to the CFT vacuum was taken to be that of the higher spin vacuum. This is true if the vacuum state has a non-vanishing chemical potential.  
One might as well consider that the global $AdS$ is the vacuum. 
The relative entropy of the states in CFT corresponding to their appropriate bulk configurations can then be calculated with respect to the $AdS$ as the reference point or the ground state. 

The leading behaviours in the small sub-system regime then reveal the following. The difference in the entanglement entropies and modular Hamiltonians between that of the higher spin black hole and global $AdS_3$ is 
\begin{align}
(& \H_{BH} - \H_{AdS})\Big|_{\text{to $\phi^2$}}= (S_{BH} - S_{AdS})\Big|_{\text{to $\phi^2$}} \nn \\
=& \ \c \Bigg[ \frac{ \left((2\pi R T)^2 + 1\right)}{72} 
+
\frac{5 }{54}  \left(\mu R\right)^2 \left(2\pi T\right)^4
+
\frac{20  }{27} \left(\mu R\right)^4 \left(2\pi T\right)^6
+
\frac{1768}{243}   \left(\mu R\right)^6 \left(2\pi T\right)^8
\nn 
\\ \nn & \qquad +
\frac{57664 }{729 } \left(\mu R\right)^8 \left(2\pi T\right)^{10}
 + \frac{668416 }{729 }  \left(\mu R\right)^{10} \left(2\pi T\right)^{12}+  \mathcal{O}\left(  \left(\mu R\right)^{12} \left(2\pi T\right)^{14}\right)   \Bigg] (\phi)^2 \  .
\end{align}
Once again, we observe a matching of $\Delta\H$ and $\Delta S$ at the leading order in interval sizes.

\def\hg{\hat{\gamma}}
\subsection{Relative entropy in the $SL(2,\mathbb{Z})$ family of solutions}
 
In this subsection we shall find the relative entropy between a smooth solution labelled by $\hat\gamma$ (discussed previously in subsection \ref{class-sol})   in the $SL(2,\mathbb{Z})$ family and the higher spin vacuum.   The relative entropy shall then give a measure of quantum distinguishability of the solution $\hat{\gamma}$ from global $AdS_3$ belonging to the $\sl2$ family. Here
\begin{align}
\hat{\gamma}= \begin{pmatrix}
a &b \\
c &d 
\end{pmatrix}  \ , \quad \hat{\gamma} \in PSL(2,\mathbb{Z}) .
\end{align}
We also define 
\begin{align}
\xi = c\tau + d .
\end{align}
The entanglement entropy of solutions corresponding the $\sl2$ family ($S_{\hg}$) is given in equation \eqref{ee-sl2}. The variation $\Delta S_{\sl2}$ between two states is then given by 
\begin{align}\label{delS-sl2}
\Delta S_\sl2 (\phi)=  S_{\hat{\gamma}}  (\phi)-S_{\vac} (\phi) .
\end{align}
For the case of zero spin-3 chemical potential we have (for $R=1$)
\begin{align}
\Delta S (\phi) \Big|_{\mu=0} =  \frac{\c}{3}\log \Bigg|\xi \ \frac{\sin\left(\frac{ \phi}{\xi} \right)}{\sin\left(\phi \right)}   \Bigg| .
\end{align}
For non-zero $\mu$, equation \eqref{delS-sl2} has the following behaviour in the regime of small $\phi \ll |\xi|$.
	\begin{align}\label{sl2-dS}
	\Delta S_\sl2 (\phi) = \frac{\c}{72 }\Bigg[ \frac{1}{\xi^2} f\left(\frac{\mu}{\xi}\right)	-  f\left({\mu}\right)	\Bigg]\phi^2 + \order(\phi^4),
	\end{align} 
where
\begin{align}
f(x) =1-\frac{20 }{3}x^2+\frac{160}{3}x^4-\frac{14144 }{27}x^6+\frac{461312 }{81}x^8-\frac{5347328 }{81}x^{10} +\order(x^{12}).
\end{align}

The change in the modular Hamiltonian, $\Delta \langle H\rangle = \langle H_{\hg} \rangle -\langle  H_{\vcm}\rangle $ can be calculated using the equation \eqref{mod-formula-3} 
\begin{align}\label{sl2-dH}
\Delta \H_\sl2 = 8 \pi  R^2 \left[1-  \frac{\phi }{2} \cot \left(\frac{\phi }{2}\right)\right] \Delta \cL _\sl2 \  ,
\end{align}
where $\Delta\cL_\sl2 = \cL_{\hg}-\cL_{\vcm}$. $\cL_{\hg}$ can be read out from \eqref{sl2z-l} and $\cL_\vcm$ from \eqref{hscs-l}. The short distance behaviour of $\Delta \H_\sl2$ does agree with that of $\Delta S_\sl2$ in \eqref{sl2-dS}. 

More generally, for sub-system sizes not necessarily short, the relative entropy between a state (corresponding to the smooth solution $\rho_{\hg_1}$ in the $SL(2,\mathbb{Z})$ family) and the vacuum in the dual CFT is 
\begin{align}
\S (\rho_{\hg}|| \rho_{\vcm}) =   \Delta \H_\sl2 - \Delta S_\sl2 \ ,
\end{align}
where, $\Delta S_\sl2 $  and $\Delta \H_\sl2$ are given by the equations \eqref{delS-sl2} and \eqref{sl2-dH}. 

%


\section{Conclusions}
\label{conclude}

The aim of this work was to verify the validity of the first law of entanglement, $\Delta \H=\Delta S$,  in the regime of sufficiently short intervals and at a finite chemical potential. This ensures the vanishing of the relative entropy for small subsystem sizes which is expected to hold true for any quantum mechanical system. The systems we were interested in were CFTs  dual to higher spin theories of gravity in the bulk. 

It is important to note, that the relative entropy was calculated between two non-perturbative states in the CFT Hilbert space. These correspond to two distinct instanton-like states in the dual gravity theory. Our calculations therefore provide a strong support in favour of the holomorphic Wilson line proposal  for  entanglement entropy in higher spin holography. The relative entropy   which we have investigated here, is independent of the way we introduce the UV cut-off (and this is special to 2$d$ CFTs). It is therefore a refined quantity in this sense when compared to entanglement entropy. 

We have also calculated the relative entropy between a member in the $\sl2$ family of smooth solutions and the vacuum. Although most solutions are sub-dominant in the calculation  of the full gravity partition, they do have a non-zero contribution to the same. The relative entropy acts as a useful measure of quantum distinguishability of the states from the dual CFT point of view and may shed some light on the Hilbert space of the theory. Even the results with  $\mu=0$ are relevant when one considers just Einstein gravity in $AdS_3$. 

We end with a short discussion on the possible generalizations of this work and some future directions. Our analysis was performed in the $SL(3,\mathbb{R})$ Chern-Simons theory in the bulk and can hopefully be generalized to the case of hs$[\lambda]$. This shall enable us to verify the validity of $\Delta S = \Delta \H$ for the dual $\cW_\infty[\lambda]$ CFT about which a lot is known. However, technical and conceptual obstacles exist at present to evaluate the Wilson line functional in the infinite dimensional representation of hs$[\lambda]$. 

It was speculated in \cite{Blanco:2013joa}, that the equality $\Delta \H =\Delta S$ presents an exciting possibility of reconstructing the modular Hamiltonian from the holographic entanglement entropy and one might potentially derive the reduced density matrix from it. It shall be fascinating to explore how such a possibility can be realized for 2$d$ CFTs with $\cW$-algebra symmetries from their holographic duals.

\section*{Acknowledgements}  
The author is grateful to Justin David for fruitful discussions, suggestions on the manuscript and encouragement. He thanks Michael Ferlaino, S.~Prem Kumar and Aninda Sinha for comments on the manuscript and discussions.  He also thanks Rajesh Gopakumar, Lin-Yan Hung, Apratim Kaviraj, Apoorva Patel, U.~Satya Sainadh, Arunabha Saha, Shaon Sahoo and Kallol Sen for discussions. 

\bibliography{entropy}
\bibliographystyle{bibstyle}
\end{document}